# DESIGN AND IMPLEMENTATION OF A WIRELESS SENSOR AND ACTUATOR NETWORK FOR ENERGY MEASUREMENT AND CONTROL AT HOME

Edwin Chobot, Daniel Newby, Renee Chandler, Nusaybah Abu-Mulaweh, Chao Chen, Carlos Pomalaza-Ráez

Department of Engineering, Indiana University – Purdue University Fort Wayne, Indiana, USA

## *ABSTRACT*

*This paper describes the design, implementation, and testing of a wireless sensor and actuator network for monitoring the energy use of electric appliances in a home environment. The network includes energy measurement nodes and a central server, where the nodes read the energy use of connected appliance, and wirelessly report their readings to the central server for processing. The server displays the readings from these nodes via a user visual interface in real time. Through this system, users can easily understand their electricity usage patterns and adapt their behaviour to reduce their energy consumption and costs. Moreover, users are able to remotely power on/off individual devices to actively control the power use of certain appliances. The system allows for inexpensive monitoring of home energy use and illustrates a practical way to control the energy consumption through user interaction.*

## 1. INTRODUCTION

In a world of rising energy costs and dwindling natural resources capable of producing energy, people and businesses are starting to look for better ways to help reduce their increasing electric bills. One way of reducing these costs is to monitor, in real time, how much power is being consumed and from these data make informed decisions about how to manage the electrical devices being powered. A system that can give users an estimate of how much energy is being, has been, and might be consumed will allow them to adjust their habits and lower the costs.

In this work, we design, build, and test a wireless sensor and actuator network called the "*wireless energy custodian network*" for monitoring the energy use of alternating current (AC) appliances in a home environment. The measured energy use of individual appliances can be displayed through a user visual interface in real time; so that users can easily understand their electricity usage patterns and adapt their behaviour to reduce their energy consumption and costs. Moreover, users are able to remotely control the power on/off of individual devices to actively control the power use of certain appliances. The system allows for inexpensive monitoring of home energy use and illustrates a practical way to control the energy consumption through user interaction.

### 1.1. System Overview

Our proposed wireless energy custodian network has two major function modules: the *energy measurement module* and the *remote power on/off control module*.

- **Energy measurement module**

The wireless sensor and actuator network consists of multiple measurement nodes and a central server, where the measurement nodes have two-way communication with the central server. Each measurement node in the network is connected to and reads the energy use of one AC appliance, and wirelessly reports the readings to the central server for processing. The server displays the readings from these nodes through a user visual interface in real time. This system can help users better understand their electricity usage patterns and adapt their behaviour to reduce their energy consumption and costs.

- **Remote power on/off control module**

Another main objective of this project is to allow users to actively control the energy usage of certain devices through the wireless sensor and actuator network. To illustrate a simple method without getting into the functionalities of specific appliances, we choose *power on/off control* in the following application scenario. Most electronic products, even if turned off, will continue to draw power from a standard electrical outlet unless the device is manually unplugged. This power consumption is called “*standby power*.” Although individual electronic products might not draw much power to be noticed while in standby mode, the average American family has almost forty devices constantly consuming power. The standby power consumption of these devices accounts for almost 10% of the whole household electricity use [1]. Because of this, our design will integrate an actuator into each measurement node that automatically turns on and off the power supply to the products remotely.

The remote on/off control can also be used in other manners to further reduce energy. For example, the air conditioner and fans can be turned on and off remotely based on the inputs from temperature sensor readings; lights can be turned on and off remotely based on the inputs from light sensor and motion sensor readings. The design of these appropriate control mechanisms, however, depends on specific appliances and the habits of individual users. Therefore, we specifically target the standby power to illustrate the feasibility and functionality of the on/off control. Furthermore, this on/off control enhances the wireless network from monitoring only to also including the actuator part, which extends the capability of the whole system and makes our design different from other commercial products on the market.

Figure 1 shows an application of this wireless energy monitoring system in a home scenario, where the measurement nodes are connected to major home appliances in different rooms, and the central server displays the energy consumption of these appliances on a computer screen.

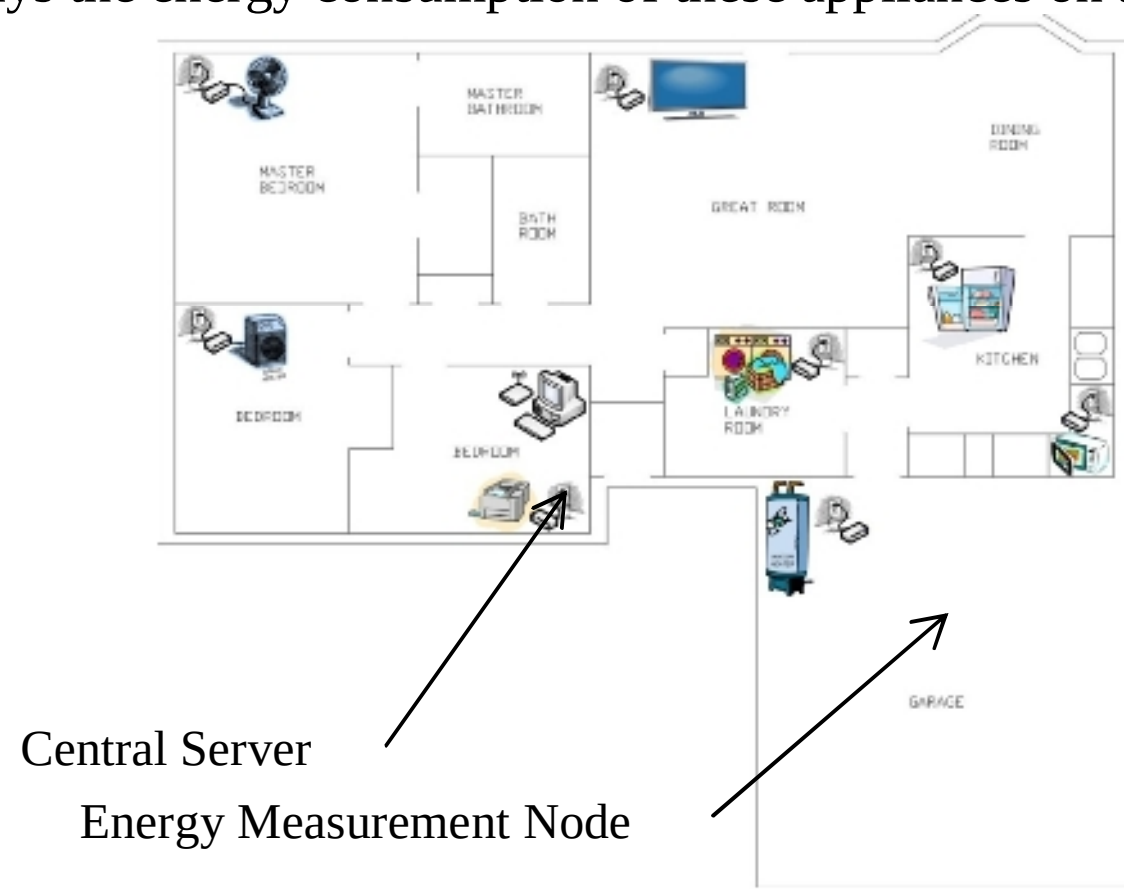

Figure 1. A wireless energy monitoring system at home.

Considering the constraints of time and skills of the undergraduate students involved in the design project, we decided to design and implement a system prototype with two measurement nodes and one central server, where the nodes communicate directly to the server and the server displays the measurement results through the computer. Requirements of the system prototype include:

- The nodes monitor and wirelessly transmit the energy usage of connected AC devices.
- The central server displays the energy reading in real-time through a graphical user interface while updating once every minute.
- The central server is able to turn on and off the individual nodes. The on/off control will be tested in the application of standby power reduction.
- The system prototype needs to be deployed and tested indoors, in a typical home environment in the United States. The communication distance from the measurement nodes to the central server is within 15 meters.

### 1.2. Related Products

Currently there are several products already on the market that can be used for power monitoring purpose. For example, the Pacific Gas and Electric Company in California has started a program to replace conventional meters by the *SmartMeter* [2]. The SmartMeter records the whole house electricity consumption on an hourly basis. The main goal of the SmartMeter technology is to enables PG&E to set pricing that varies by season and time of the day, rewarding customers who shift energy use to off-peak periods. *TED (The Energy Detective)* [3] is an in-home electricity monitor that measures home energy usage through residential electric panels and updates real-time monitoring results every second. TED displays the results on an LCD screen and/or a local or remote computer using its data-logging software or using Google PowerMeter [4], a free energy monitoring software. *Kill-A-Watt* [5] is a device which monitors the amount of electric consumption of a connected appliance by the kilowatt-hour and displays it on an LCD display. Users can calculate the electrical expenses by the day, week, month, or year.

Compared with the above products on the market, our proposed energy monitoring system aims to monitor the energy consumption of individual appliances (like Kill-A-Watt) and provides real-time measurement results (like TED), so that users know where they can save more energy and are able to get immediate feedback on their power usage. Moreover, our system is capable of controlling individual appliances automatically or remotely through a computer. Although simple, this on/off control enhances the wireless network from monitoring only to including the actuator part, which extends the capability of the whole system and distinguishes our design from other products on the market. If the power on/off control is feasible, the system can be easily extended to incorporate other more complicated control mechanisms. Due to the limitations in time and resources, our senior design project will only focus on a power on/off control mechanism.

## 2. DETAILED COMPONENT DESIGN

Our system prototype has two measurement nodes and a central server. Each measurement node will be plugged into a standard NEMA 5-15 electrical outlet. An AC device will then be plugged into the node for power measurement. Each measurement node contains the components necessary to measure the power consumption, wirelessly transmit the information to the central server, and control the power on/off of the connected device. The layout of the functional blocks of the measurement node is shown in Figure 2. The energy measurement block monitors the voltage and current that are passed to the connected device, and calculates the power consumption. The microcontroller then reads the power measurement signal and reports to the server through the wireless transceiver. The microcontroller also has the capability of controlling the power delivered to the connected device through the actuator block. The AC-DC power supply block converts the AC voltage source to fixed DC voltage sources that the energy measurement block, the microcontroller, and the wireless transceiver need to operate on. In addition, the status indicator block identifies the node's current mode of operation, and a reset button allows the user to locally restore power to a device that is currently off.

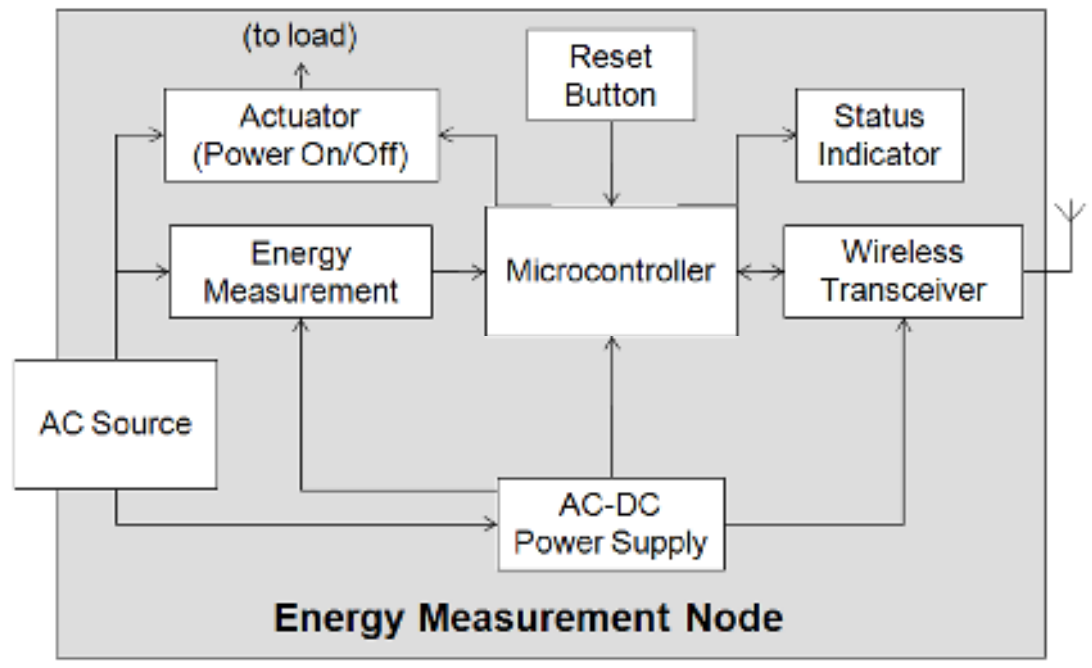

Figure 2. The block diagram of the measurement node.

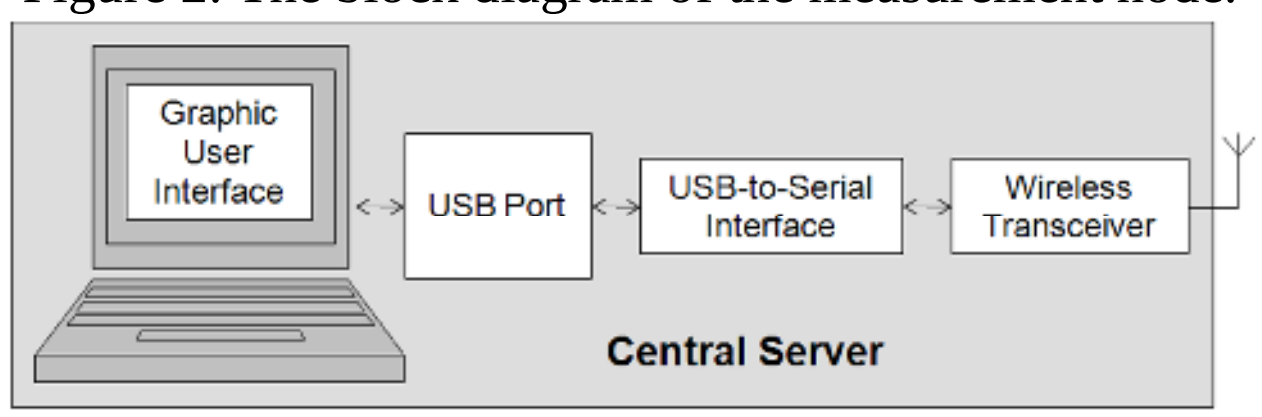

Figure 3. The block diagram of the central server.

The central server receives the power measurements from each of the nodes and forwards the measurements to the computer for display. The measurement data is received through a wireless transceiver and passed directly to the computer program through the USB port. The power on/off signal from the computer program is passed to the wireless transceiver through the USB port. The components in the central server are powered through the USB port as well. Figure 3 shows a block diagram of the main components included in the central server.

In the following parts of this section, we give the detailed design of the major components in the measurement nodes and the central server.

## 2.1. AC-to-DC Power Supply

In our design, each measurement node will be plugged into a standard NEMA 5-15 electrical outlet. An AC device will then be plugged into the measurement node. The circuitry used to measure the power will need to be powered by DC voltage. This may include several different DC voltage levels, such as 3.3V and 5.0V. The plan for our design is to tap the AC power and convert it the DC power required by the measurement node's internal circuitry.

Our AC-to-DC power supply consists of three stages. The first stage is a step-down transformer that scales down the AC voltage by a specified factor, such as 10:1. This has the benefit of allowing access to a smaller AC voltage that can be used for our voltage measurements. Additionally, the voltage output by the secondary coil of the transformer is electrically isolated from the primary coil (main AC power supply). Next, the secondary AC signals are rectified by using a two-diode full-wave rectifier. In the last stage, the voltage waveform is converted to DC voltage using a filter and voltage regulators. Figure 4 below shows the schematic of the AC-to-DC power supply that has been adopted in our design.

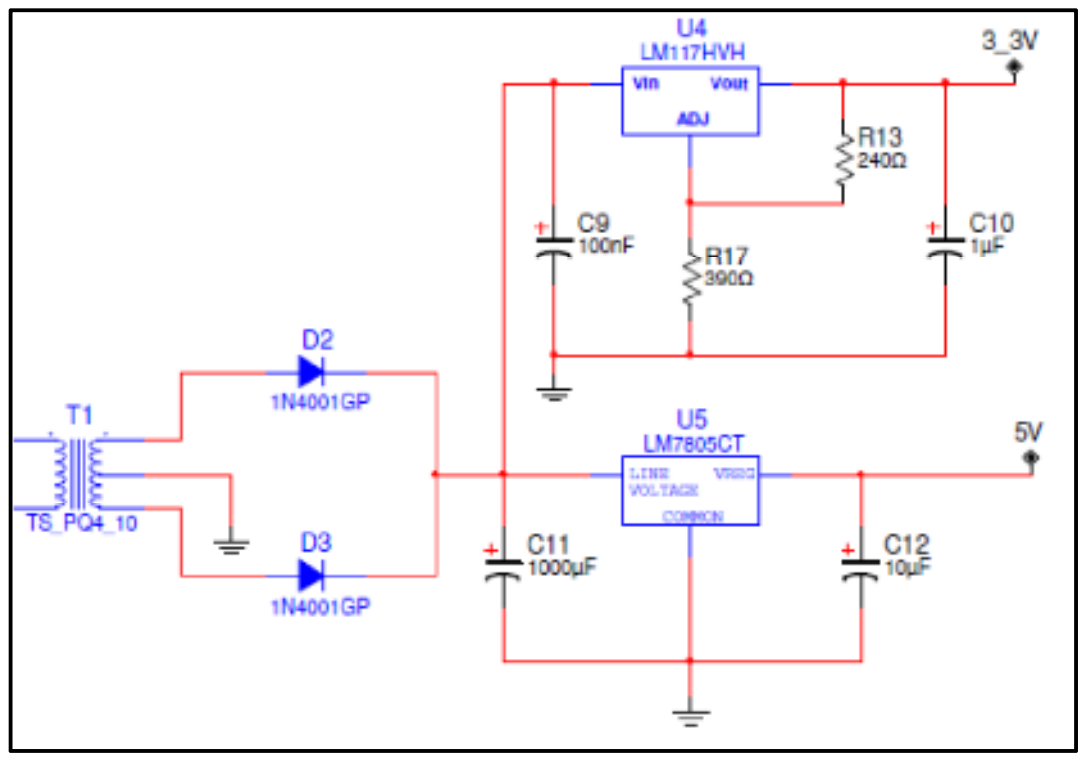

Figure 4. The schematic for the AC-DC power supply.

## 2.2. Energy Measurement

To calculate the real-time energy consumption of a connected AC device, we need to measure the input voltage and the current flowing through the device. The power supplied by a NEMA 5-15 standard electrical outlet is delivered as a sinusoidal waveform with a frequency of 60Hz. According to the Nyquist-Shannon sampling theorem, a sampling rate greater than 120Hz is required to accurately measure the voltage and current through the connected appliance. During each sample period, the analog samples will be converted into a digital signal with an analog-to-digital converter (ADC). Additionally, the voltage and the current will need to be sampled simultaneously in order to accurately calculate the instantaneous power usage. This requires that our design includes two separate ADCs.

Since ADCs require an input voltage instead of a current for their conversions, the current passing through the load appliance must be converted to a voltage before recordings can occur. After the recordings are taken, the voltage measurement will be converted to a current measurement and power calculations can take place. We considered the following two options for the current-to-voltage conversion:

- *Hall Effect Sensor:* A Hall Effect sensor is a transducer whose output voltage varies with changes in magnetic field. Changes in magnetic field occur if the current flowing through the load appliance changes. By using a Hall Effect sensor, the interference with the actual load can be kept to a minimum.
- *Low Impedance Resistor:* A low impedance resistor can be used in series with the load to determine the current flowing through the load by using Ohm's law. While this is simple, it slightly alters the voltage delivered to the load because the resistor reduces the voltage delivered to the load.

The low impedance resistor is chosen in our design since it takes up less space and is less expensive than the Hall Effect sensor. Specifically, we selected the Panasonic ERJM1WS 15mΩ resistor [6], which costs $1.18 each.

To measure the power consumption, we chose the Analog Devices AD71056 energy metering chip [7]. This chip contains two 16-bit ADCs for the voltage and current signal inputs, which provides very precise power measurements. The analog input range is fully differential (±30mV for the current signal and ±165mV for the voltage signal), which does not require adding the DC offset to the AC signals that are necessary for single-ended analog input ranges such as 0-5V. A high sampling rate of 450kHz is used, which is well above the minimum sampling rate required. There are three output signals specifying the power measurement: One high frequency output supplies instantaneous real power, and two low frequency outputs supply average real power. Each frequency output is a square-wave signal whose frequency is linearly dependent on the electrical power. This chip requires a small amount of power (20mW) and is priced at $2.63 each.
The complete energy measurement circuit is shown in Figure 5, where the purpose of the voltage signal conditioning is to scale down the AC voltage signal to the specified input range of ±165mV. The AC current signal is converted to a voltage signal through the low-impedance sense resistor (R11 in Figure 5) and the current signal conditioning limits the range of the input range for this signal to ±30mV. Both signals are band limited with a low-pass RC filter with 3dB cut-off frequency at 5.3kHz. This will reduce any aliasing of the signals when sampled by the ADCs. In addition, the attenuation to the 60Hz input signals is negligible. The optocoupler provides the electrical isolation and the voltage conversion between the square-wave output signal from the energy metering chip and the microcontroller. The microcontroller then reads in and converts the square-wave signal to a digital word.

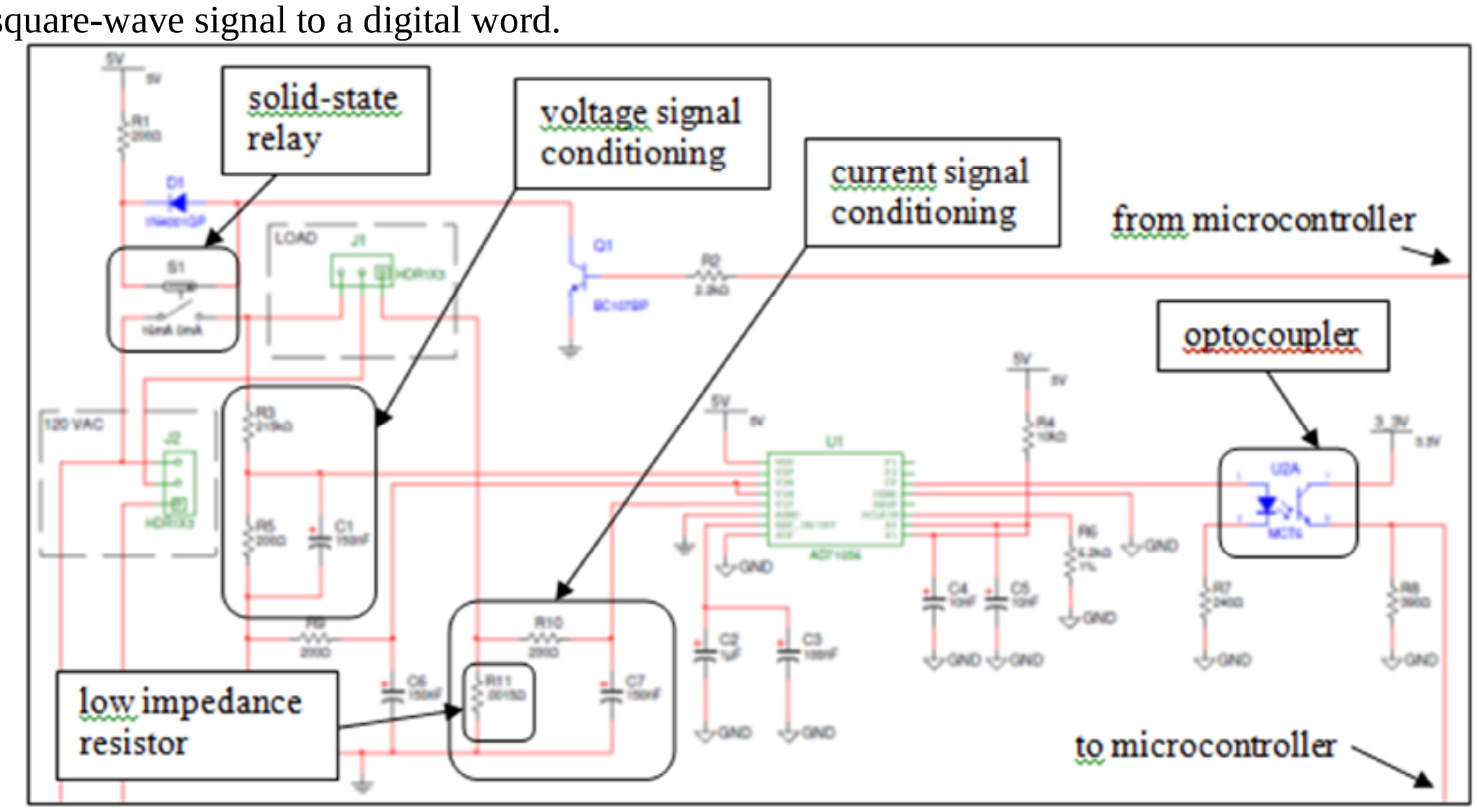

Figure 5. The schematic for the energy measurement and the actuator circuit.

## 2.3. Actuator

An actuator will be placed in each node with which the user can turn on or off the current flowing through the load. We choose a solid state relay (S1 in Figure 5) for this purpose. The specific part we picked is the Sharp Microelectronics S216S02F solid state relay [8], which costs $6.45 each.

## 2.4. Wireless Transceiver

Our system is targeted for indoor use at a typical American home; therefore a short-distance wireless communication system is more appropriate. Two types of wireless communication standards are suitable for this application: IEEE 802.15.1 (Bluetooth) for medium rate wireless personal area networks (WPAN) and IEEE 802.15.4 (ZigBee) for low rate WPANs, both operating in the 2.4GHz unlicensed industrial, scientific, and medical (ISM) frequency band.
IEEE 802.15.1 is adapted from Bluetooth, which specifies short-range RF-based connectivity for portable devices. Bluetooth is designed for small and low cost devices with low power consumption. Since Bluetooth is geared towards handling voice, images, and file transfer, it has a data transfer rate on the order of 1Mbps with a relatively complex protocol. The operational range for Bluetooth is around 10 meters. With amplifier antennas its range can be boosted to 100 meters, but with higher power consumption.

IEEE 802.15.4 handles low-cost, low-speed ubiquitous communication between devices. It is designed for equipments that need a battery life as long as several months to several years but do not require a data transfer rate as high as those enabled by Bluetooth. The 802.15.4 compliant devices have a transmission range between 10 and 75 meters and a data transfer rate of 250kbps (if operating at 2.4GHz frequency band). 802.15.4 supports a basic master-slave configuration suited to static star networks of many infrequently used devices that talk via small data packets. Compared with Bluetooth, 802.15.4 is more power-efficient because of its small packet size, reduced transceiver duty cycle, reduced complexity, and strict power management mechanisms such as power-down and sleep mode.

For our design, we chose the IEEE 802.15.4 standard, and specifically, the XBee 802.15.4 OEM RF module [9] from DigiKey as the wireless transceiver. It can connect directly to a microcontroller with a UART interface. With a chip antenna, it operates up to 30 meters indoor. The transmission range can be further increased to 90 meters by using a whip antenna. The XBee module has a low maximum transmit power of 1mW and a high receiver sensitivity of -92dBm. Each transceiver can be uniquely identified by its media access control (MAC) address. The XBee transceiver is priced at $19.00 each.

## 2.5. Status Indicator

The measurement nodes will have some visual indicators to alert the user of the status of the node. We chose three LEDs with different colours for the following indications:

- Green: The measurement node is functional.
- Yellow: The measurement node is transmitting data to the server.
- Red: The power to the load device has been shut off.

## 2.6. Microcontroller

The measurement nodes need a microcontroller to do simple processing on the sampled power measurement data from the energy metering chip, turn on or off the solid state relay, and control the flow of the data and control message. We chose Silicon Labs C8051F353 microcontroller [10] with a built-in 16/24 bit ADC in our design. The cost of this microcontroller is $7.20.

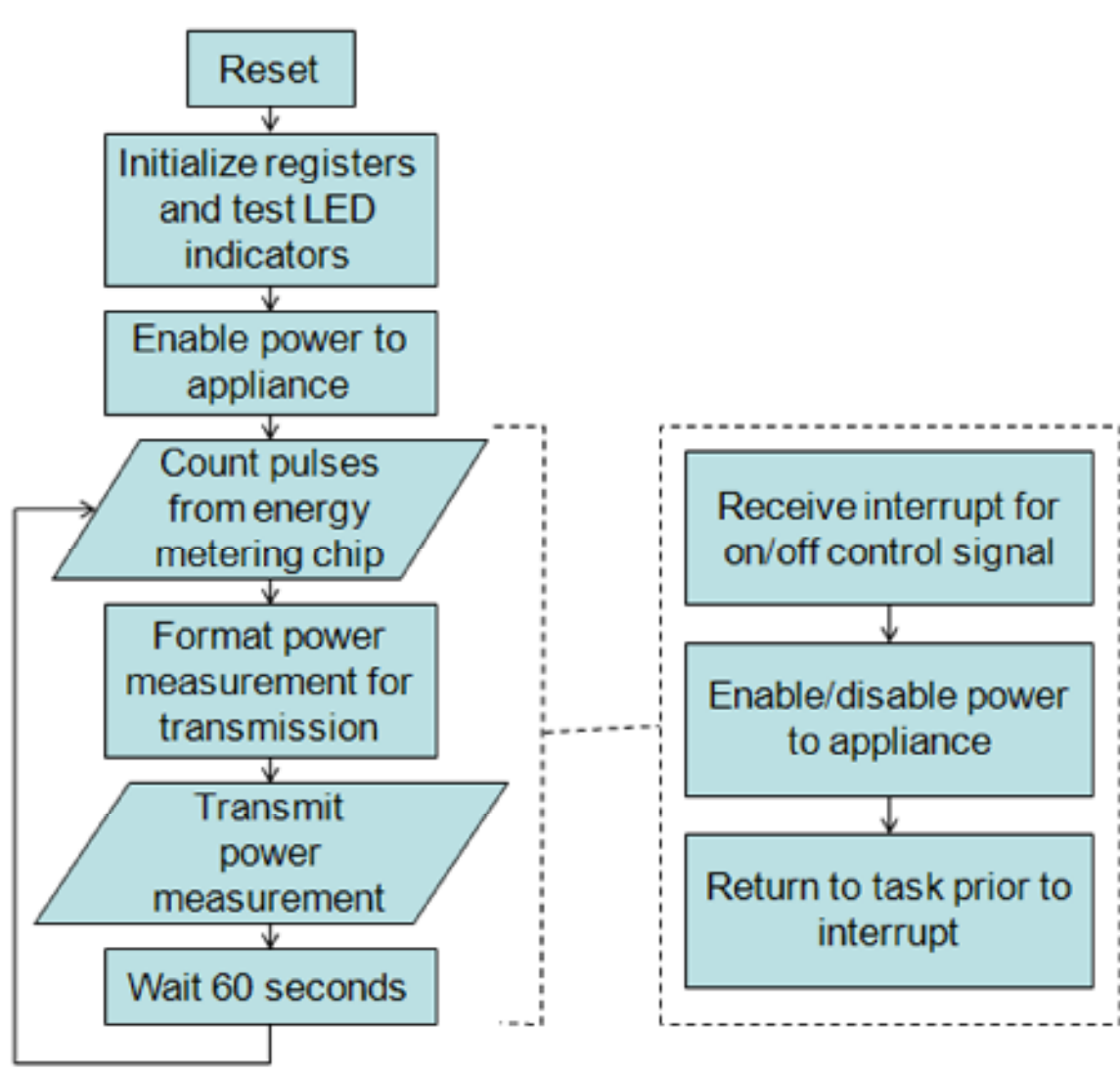

Figure 6. Flowchart for the measurement node software.

The main function of the software is to collect measurements from the energy-metering chip, transmit the measurements to the central server, receive control signals from the server, and control the power supplied to the connected appliance. We use two bytes for the frequency count data, resulting in a power resolution of about 0.02 watts. The flowchart in Figure 6 illustrates the operation of the node measurement software.

## 2.7. Central Server

To match the wireless transceiver in the measurement node, we used the same XBee transceiver in the central server. Specifically, we selected the XBee Explorer Dongle unit [11] from the SparkFun Electronics. This unit has a USB UART interface IC device (FT232RL) that interfaces between the XBee transceiver and the USB port. It also equipped with a USB port and a dock for the XBee transceiver. The XBee Explorer Dongle unit costs for $24.95.

The graphical user interface (GUI) in the central server displays the energy usage measurement in real-time and provides for the control of the power on/off of the appliances through the measurement nodes. The design of the GUI software will be described in detail in Section 3.1.

## 2.8. Design Summary

Figures 7 and 8 show the photographs of an assembled measurement node and the central server, respectively. The overall size of the measurement node is 12cm×12cm×6cm. In our system prototype, the total cost for each measurement node is $112.79 (including $30 for PCB board manufacturing) and the total cost of the central server is $43.95. If hardware components can be purchased in larger quantities (e.g., 1000 units), the cost of a measurement node will be reduced to $59.12.

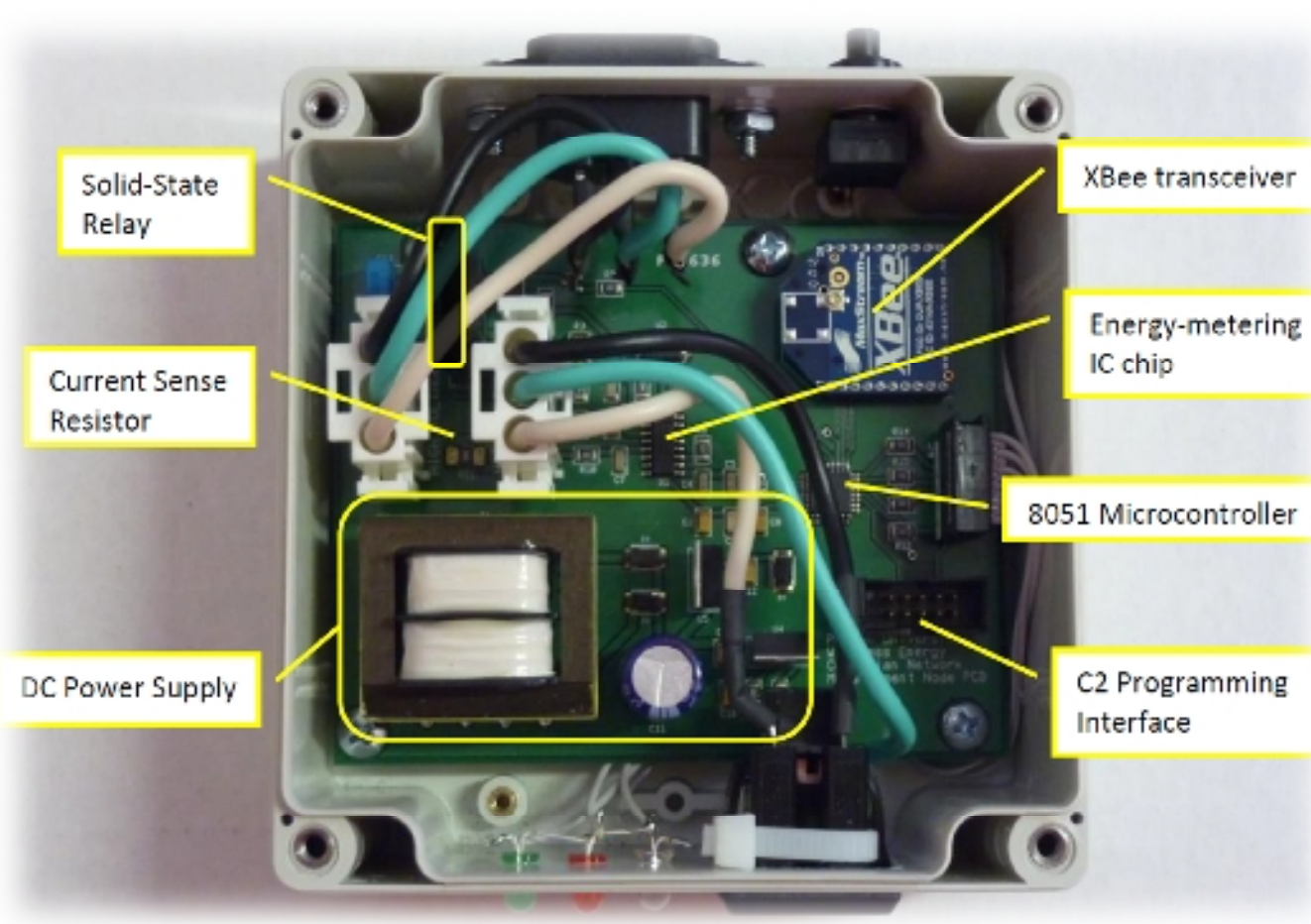

Figure 7. Photograph of the completed measurement node.

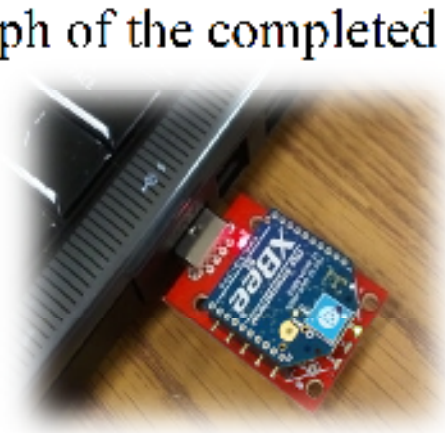

Figure 8. Photograph of the central server.

## 3. SYSTEM TESTING AND RESULTS

For our design we tested each of the main components required for proper operation of the nodes and the central server. This included measuring the power consumption at the node, wirelessly transmitting the information to the central server, rendering the information on the graphical user interface (GUI) in real time, and transmitting a control signal back to the node if necessary. For the measurement nodes, the individual components tested included the AC-DC power supply, the energy measurement circuit, the reset button and the solid state relay, the XBee transceiver, and the LED status indicator. We also used the Kill-A-Watt monitor to measure the power consumption of the measurement node and found that the node currently consumes about 1.5W while transmitting data. For the central server, the components tested included the XBee transceiver and the GUI software. The GUI software design and evaluation is discussed in further detail below.

### 3.1. GUI Design and Evaluation

The GUI software was written for a Windows based personal computer and tested using the Microsoft Visual Studio development suite. The main purpose of the GUI software is to display the power consumption data in real time and control the solid-state relay for each of the nodes. To achieve these goals, the software is organized into two threads: *the data transmission thread* and *the user input thread*. The data transmission thread will execute once per minute, as that is the rate at which the nodes transmit their data. The data transmission thread consists of opening a specified serial port, reading the wirelessly transmitted data, adding a timestamp, checking the standby power requirements, saving the power measurements to a .csv file, and then updating the graphical display with the latest power measurement. Figure 9 illustrates the flowchart of the data transmission thread.

The user input thread allows the software to asynchronously handle input from the user each time the user chooses a menu item on the GUI. There are five pull-down menus: Menu, CSM, Monitor, Node, and Display.

- "Menu" has one submenu "Exit" that allows the user to confirm exit the GUI program.
- "CSM" allows the user to select the USB port to the connect server device.
- "Monitor" enables the user to start or stop the data transmission thread.
- "Node" provides the backward control function, which enables the user to enable/disable the power save feature to a specific node or send enable/disable power control signal to a specific node.
- "Display" allows the user to clear historic data or calculate total energy consumption over specified time duration.

Figure 10 illustrates the testing of a space heater with only the fan on (no heat) and an LCD computer monitor. Each of these two devices is plugged into a node and their power consumption are reported to and displayed at the GUI software once every minute.

The GUI software also has the capability to calculate the total energy used over a specified period. The user can provide a "start date" and "end date" for the software to calculate the total energy consumed by both nodes. To calculate the energy, the software reads in historical data from the .csv file and multiplies the time duration values by the power usage value. Then the data are summed up and converted from Joules to kilowatt-hours and then displayed to the user. Figure 11 shows the window displaying the energy consumption for a user-specified time period.

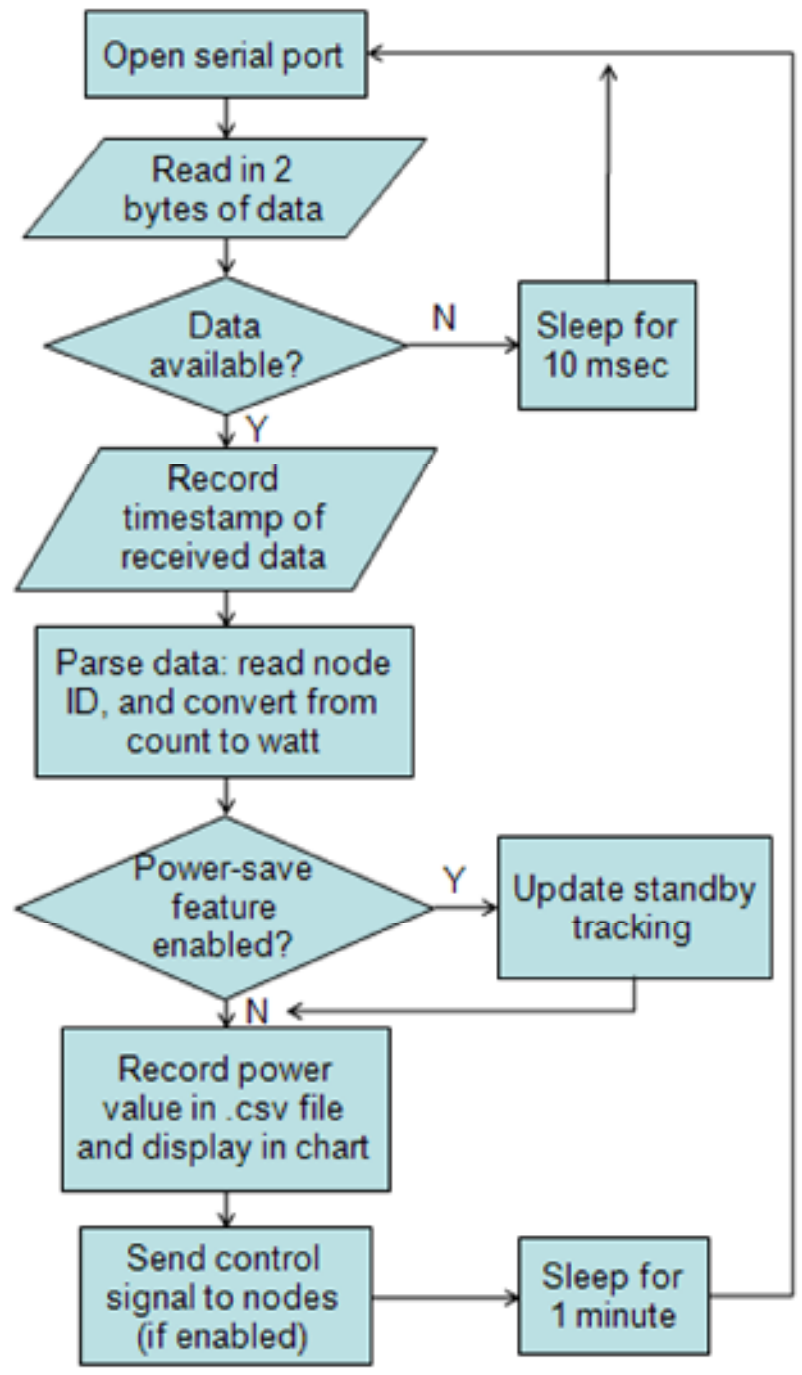

Figure 9. Flowchart for the data transmission thread.

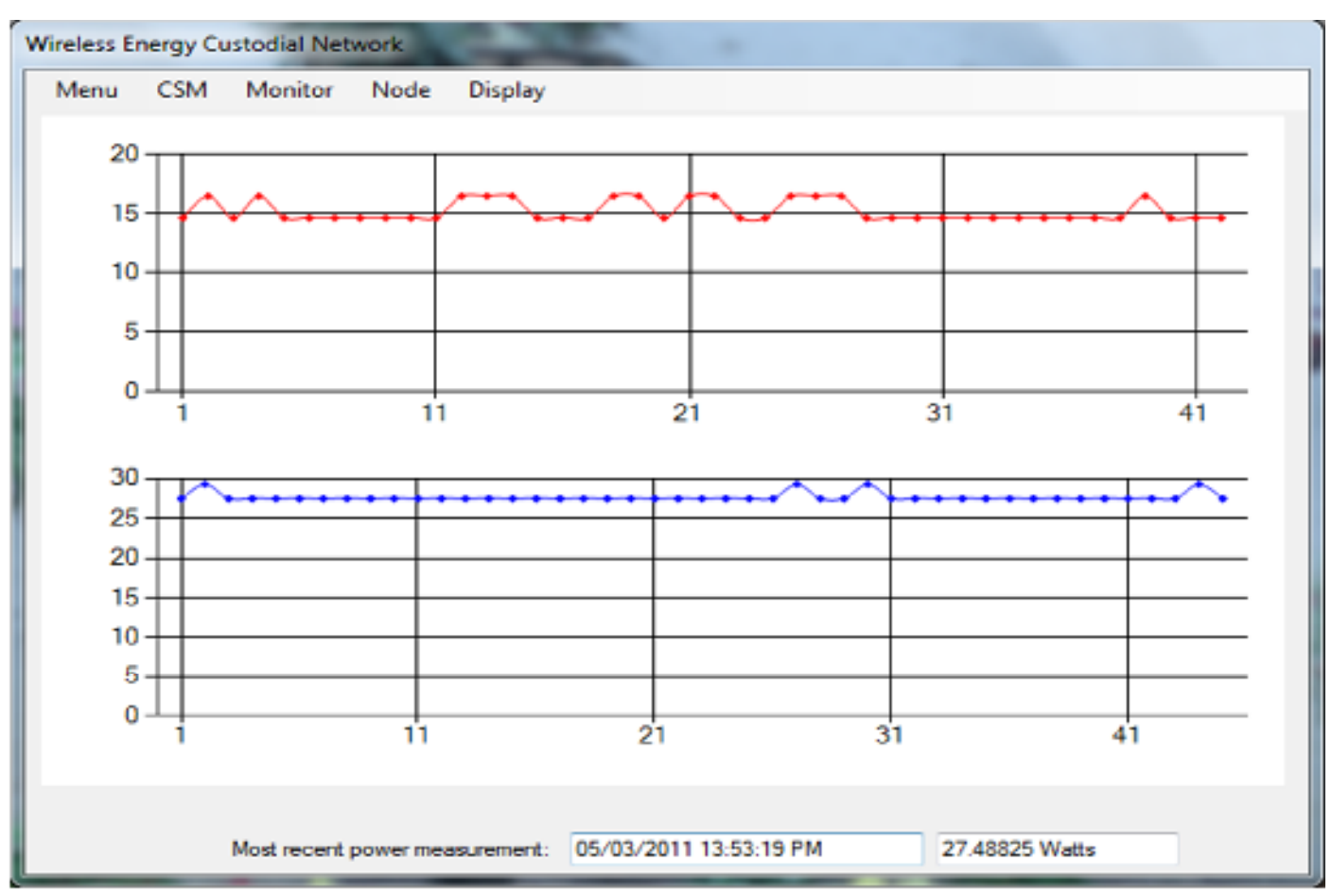

Figure 10. Screenshot of GUI displaying the power consumption for two nodes (top: a space heater with only the fan on; bottom: an LCD computer monitor).

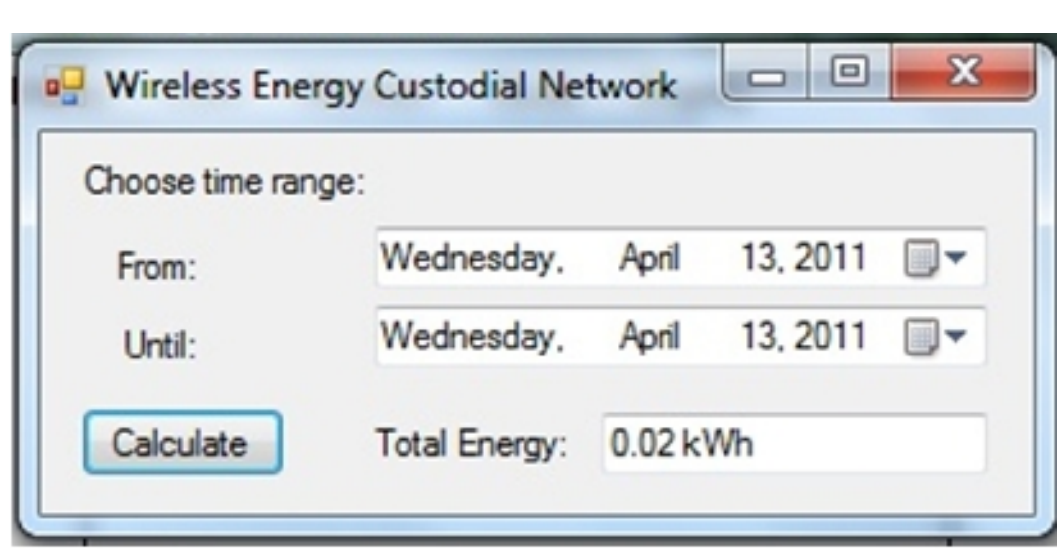

Figure 11. Calculating energy consumption over a time period.

An additional power-save feature is included in the GUI software. The purpose behind the power-save feature is to reduce power consumption by devices that go into a "standby" mode. When the standby power-save feature for a specified node is enabled, the device that is connected to that node is first put to the standby mode, and the GUI software gathers the next 10 readings it receives for that node and averages them. The average is multiplied by 120% and set as the threshold for that node. Then the device is set to normal working condition. Once the device has operated for 30 consecutive minutes (i.e., 30 consecutive readings) below the threshold, the GUI software then determines that the device is working under the standby mode and shuts off power to the device via the wireless control signal. Any reading above the threshold resets the count of "below threshold" values.

In order to test the standby power-save feature, we monitored a computer while it was consuming standby power. We first enabled the power-save feature for the computer node, and then waited for the 10 samples to be calculated in the threshold calculation phase. After the standby threshold was calculated, we turned on the computer to show usage above the threshold, which should not shut power off to the device. Then, we put the computer back into standby mode, of which the power consumption is below the threshold. After 5 consecutive samples under the threshold were received, power was automatically shut off to the computer. Here we used 5 consecutive samples instead of 30 for the purpose of reducing the testing time and showing a live demo. Figure 12 shows the GUI after the standby power-save feature was enabled for the computer. As indicated by Figure 12, the computer still consumes around 20W (about 27% to 33% of its power in normal working condition of 60-75W) while in standby mode. The GUI software was able to remotely turn off its power automatically after the computer has been in standby mode for a while. Therefore, our designed system is able to help users reduce their energy consumption by reducing the standby power consumption. Once the power of the device is cut off by the server, a user can restore the solid-state relay by pressing a push-button at the node, so that the device can be turned on again locally.

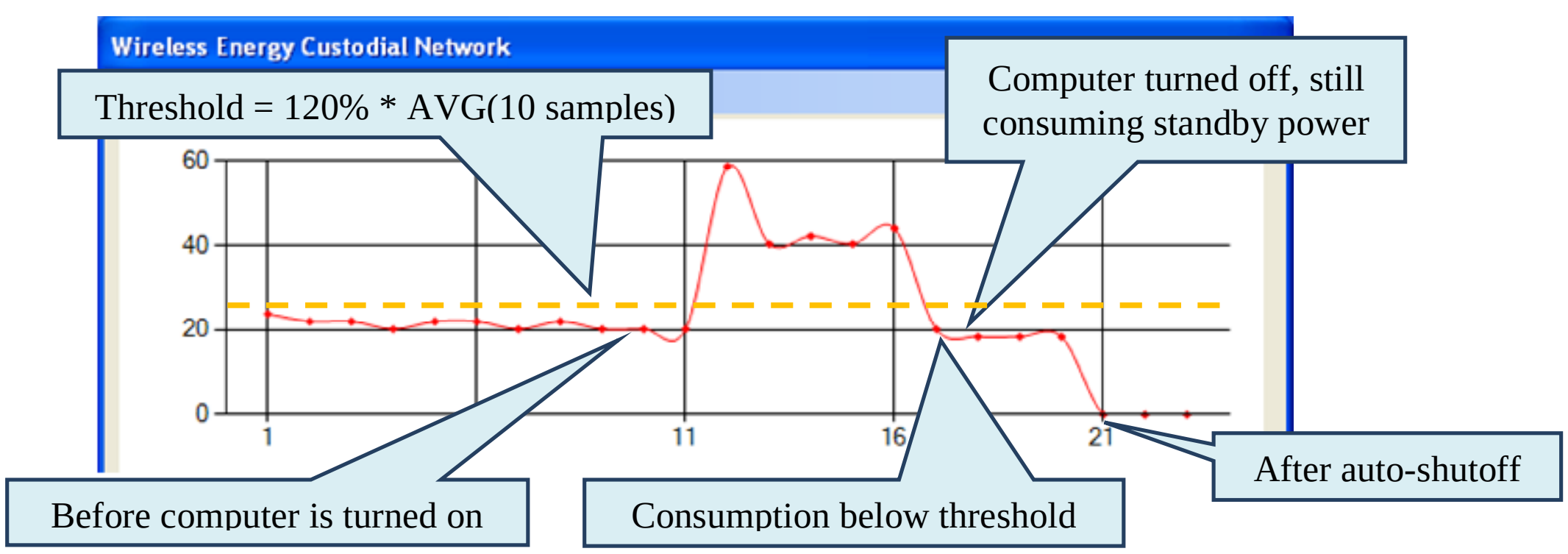

Figure 12. Screenshot of GUI displaying the standby power-save feature.

## 4. SYSTEM ANALYSIS AND DISCUSSIONS

In this section, we discuss several important aspects and possible improvements of this system prototype design, including component selection, real-time requirement, power on/off control, GUI software extension, and system expansion.

### 4.1. Component Selection

Several factors were considered when selecting hardware components. The first consideration was the selection of the component package to use for each part. Most of the electrical components in our design were selected in a surface mount package. This reduced the foot print required for each component and the overall space required on the printed circuit board. Several through-hole components were used when a comparable surface mount component was not available. Another consideration was the tolerance required for each component. Several resistors with a tolerance of ±1% were required in the circuit when accuracy was critical. Additionally, the power and voltage rating for each of the components was considered to make sure that none of the components would become overstressed during use. Finally, cost and availability were considered. By using standard components, we were able to keep our cost low and we were also able to purchase components in the low quantities required for our project.

We recommend making the following changes on the hardware components of the prototype measurement node after running into some problems during testing:

- *Solid state relay:* During testing, we discovered that running an appliance that consumes large amounts of power can make the solid state relay overheat, which may lead to product failure if the connected device is left on for too long. A solution to this would be to add a heat sink to the solid state relay. This relay has space for a heat sink to be added which would keep the relay below a safe temperature during extended periods of high power consumption.
- *Low impedance resistor:* We tested the power measurement converted from the output frequency of the energy metering chip for several different electrical devices. We also used a commercial Kill-A-Watt monitor to measure the power consumption of the same devices and used the new measurement as a reference. We found that the voltage across the current sensing inputs to the energy metering chip is larger than expected. After further analysis, we discovered that the error was caused by the additional resistance (about 0.7mΩ) through the printed circuit board track, which was ignored during the design process. This error can be corrected by using a copper track with 1.5mΩ of resistance in place of a low impedance resistor.

## 4.2. Real-Time Requirement

Generally, to allow the users to view their energy use in real-time, it is sufficient to update the GUI data every minute. The data report rate from the measurement nodes can be set to the same rate with some retransmission mechanism under data losses. However, this data report and update rate can be adjusted according to the electrical device that a node is connected to. For example, the data report rate can be set to high when the microwave is on and low when the microwave is off. There are two advantages of this adaptive report rate control: First, the nodes can switch to a lower transmission frequency to reduce the power consumption. Second, there will be fewer packets transmitted over the wireless medium, which leads to less collision and higher accuracy of the whole system.

## 4.3. Power On/Off Control

When we tested the remote power on/off control for standby power reduction, we first let the device go to the standby mode and calculated the standby threshold using power measurement samples during this period. In real-life application, the server software can be modified to detect the standby mode and set the threshold accordingly on the run. This can improve the ease of use of the system for non-technical users. In addition, the control method can also be time-based, where a user sets certain time frames that he/she wants to cut off the power of some devices. This would eliminate the standby power consumption in those nodes during those time frames.

### 4.4. GUI Software Extensions

The GUI software at the central server can have at least the following extensions:

- *History graph*: Past energy consumption data can be displayed using histograms of hourly, daily, weekly, and monthly options. This functionality can be added to future versions of the GUI software to enhance the number of options for the user.
- *Database integration*: The data can be stored and retrieved at a faster speed, if a database system is integrated with the GUI software.
- *User-defined energy price*: Allowing users to input the energy price enables the user to roughly estimate their electric bill in terms of monetary cost instead of kilowatt hour.

### 4.5. System Expansion

Our system prototype has two measurement nodes and we use two bytes for power measurement data transmission. IEEE 802.15.4 standard supports star topologies with at most 254 units. Therefore, more measurement nodes can be added into the system. To accommodate more measurement nodes, the data format can be expanded to add more bytes. Moreover, longer data packets can support higher data resolution as well.

On the other hand, if more measurement nodes are added into the system, the data reporting of the nodes needs to be coordinated to avoid or reduce signal collision. The following options can be implemented:

- The nodes transmit the results whenever they are ready. This may cause a collision. Therefore, some collision avoidance mechanism may be needed.
- The central server polls the individual nodes for measurement results. This can avoid a collision but it adds the transmission overhead.

The use of the contention free period (CFP) designed for real-time applications in the IEEE 802.15.4 medium access control protocol can also be explored to evaluate the data latency in such a system. Furthermore, other wireless signals in the house, if operating in the same frequency band, may interfere with the wireless signal transmissions of the system [12,13]. The effect of these signal interference (*e.g.*, WiFi signals) on the performance of the system needs to be tested as well.

In addition to using a computer device to monitor and control the energy usage, if the server is connected to a public network such as the Internet or a cellular network, users can remotely oversee their home electrical power usage even when they are away from home.

## 5. RELATED WORK

With increased awareness in energy conservation, companies and research groups have attempted to design power monitoring solutions for residential and commercial use. In addition to the related commercial products mentioned in Section 1, several power monitoring and control systems have been proposed and implemented in prototype forms at university research labs.

Different electric power meter modules have been designed based on various selections of energy metering IC chips and current transform circuit [14,15,16]. We referred to these designs in constructing our energy measurement module. A self-powered MEMS sensor module for measuring and reporting instantaneous power is described [17]. In this design, the on-board MEMS sensors are passive proximity sensors that do not require galvanic contact with the energized conductor. Its energy of operation is supported by an AC energy scavenger that harvests the energy from the fluctuations of the magnetic field in the electrical circuit it is monitoring. This design is non-intrusive and energy efficient for power monitoring. However, the energy harvested in this way may not be enough to support the actuator functions such as the remote on/off control in our design.

The communication technologies that are used to transfer the measured power information from the power meter to a control or management device vary with different designs. These technologies can be classified into wireline and wireless. Wireline transmission can occur through serial, USB, or Ethernet cables to a connected computer [14], or even through power line communication [18]. Wireless technologies such as wireless LAN, Bluetooth, and ZigBee all have been adopted [15,19,20]. Our system uses ZigBee that has low power consumption, a medium data transfer rate, and is easy to expand and fit for use in typical home scenarios.

Besides continuous power monitoring of home appliances, other control functions have been considered in existing system design as well. For instance, an existing design enables automatic recognition and malfunction detection of certain major home appliances [14]. A solid-state relay is added as an actuator in a power management system to cut off the power if the connected device is overloaded [15]. An energy control system has a redesigned power outlet that shuts off the power in standby state [20]. A proposed energy management unit communicates with the home appliances, smart meter, and power storage units, and schedules appliances to work in less expensive hours to minimize the energy expenses of the consumers [21]. Our system has remote on/off control that is able to turn on/off the power supply to the connected device remotely. We have demonstrated the capability of standby power reduction in our prototype design. Nevertheless, it is easy to expand the system to incorporate other remote control functionalities.

## 6. CONCLUSION

This paper describes a design project that builds a wireless sensor and actuator network for monitoring the energy usage of AC appliances in a home environment. The design of the system prototype including two measurement nodes and a central server is explained. The system prototype meets the design criteria. Additionally, the implementation and performance analysis of this design project have been completed. The system design and implementation illustrates an inexpensive way to monitor the home electrical energy use and control the energy consumption through user interaction.

## REFERENCES

[1] *Standby power*. Retrieved October 2012, from http://standby.lbl.gov/standby.html.

[2] *SmartMeter™ - See your power*. Retrieved October 2012, from http://www.pge.com/myhome/customerservice/smartmeter.

[3] *The Energy Detective*. Retrieved October 2012, from http://www.theenergydetective.com/.

[4] *Google PowerMeter – Save energy. Save money. Make a difference.* Retrieved October 2012, from http://www.google.com/powermeter/about.

[5] *Kill A Watt*, Retrieved October 2012, from http://www.p3international.com/products/special/P4400/P4400-CE.html.

[6] *Low Resistance Value Chip Resistors (Current Sensing Resistors).* Retrieved October 2012, from http://industrial.panasonic.com/www-data/pdf/AOA0000/AOA0000CE7.pdf.

[7] *Energy Metering IC with Integrated Oscillator and Reverse Polarity Indication.* Retrieved October 2012, from http://www.analog.com/static/importedfiles/data_sheets/AD71056.pdf.

[8] *S116S02 Series/S216S02 Series, $I_T$(rms)≤16A, Zero Cross type , SIP 4pin, Triac output SSR*. Retrieved October 2012, from http://media.digikey.com/pdf/Data%20Sheets/Sharp%20PDFs/S116,216S02.pdf.

[9] *XBee® ZB ZigBee® PRO Modules*. Retrieved October 2012, from http://www.digi.com/products/wireless/zigbee-mesh/xbee-zb-module.jsp.

[10] *C8051F35x Analog-Intensive MCUs*. Retrieved October 2012, from http://www.silabs.com/products/mcu/mixed-signalmcu/Pages/C8051F35x.aspx.

[11] *XBee Explorer Dongle*, Retrieved October 2012, from http://www.sparkfun.com/products/9819.

[12] Chen, C., & Pomalaza-Ráez, C. (2010) "Implementing and Evaluating a Wireless Body Sensor System for Automated Physiological Data Acquisition at Home," *International Journal of Computer Science and Information Technology*, Vol. 2, No. 3, pp 24-38.

[13] Yi, P., Iwayemi, A., & Zhou, C. (2011) "Developing ZigBee Deployment Guideline Under WiFi Interference for Smart Grid Applications," *IEEE Transactions on Smart Grid*, Vol. 2, No. 1, pp 110-120.

[14] Serra, H., Correia, J., Gano, A. J., de Campos, A. M., & Teixeira, I. (2005) "Domestic Power Consumption Measurement and Automatic Home Appliance Detection." *2005 IEEE International Workshop on Intelligent Signal Processing*, pp. 128-132.

[15] Bai, Y.-W., & Hung, C.-H. (2008) "Remote Power On/Off Control and Current Measurement for Home Electric Outlets Based on a Low-Power Embedded Board and ZigBee Communication," *2008 IEEE International Symposium on Consumer Electronics*.

[16] Jiang, X., Dawson-Haggerty, S., Dutta, P., & Culler, D. (2009) "Design and Implementatin of a High-Fidelity AC Metering Network," *2009 International Conference on Information Processing in Sensor Networks*, pp 253-264.

[17] Paprotny, I., Leland, E., Sherman, C., White, R. M., & Wright, P. K. (2010) "Self-powered MEMS Sensor Module for Measuring Electrical Quantities in Residential, Commercial, Distribution and Transmission Power Systems," *2010 IEEE Energy Conversion Congress and Expositio*, pp. 4159-4164.

[18] Lien, C.-H., Chen, H.-C., Bai, Y.-W., & Lin, M.-B. (2008) "Power Monitoring and Control for Electric Home Appliance Based on Power Line Communication," *2008 IEEE International Instrumentation and Measurement Technology Conference* , pp. 2179-2184.

[19] Lien, C.-H., Bai, Y.-W., & Lin, M.-B. (2007) "Remote-Controllable Power Outlet System for Home Power Management," *IEEE Transactions on Consumer Electronics*, Vol. 53, No. 4, pp 1634-1641.

[20] Han, J., Lee, H., & Park, K.-R. (2009) "Remote-Controllable and Energy-Saving Room Architecture based on ZigBee Communication," *IEEE Transactions on Consumer Electronics*, Vol. 5, No. 1, pp 264-268.

[21] Erol-Kantarci, M., & Mouftah, H. T. (2011) "Wireless Sensor Networks for Cost-Efficient Residential Energy Management in the Smart Grid," *IEEE Transactions on Smart Grid*, Vol. 2, No.2, pp 314-325.